\documentclass[twocolumn,english,aps,prd,graphics,floatfix,a4paper,superscriptaddress]{revtex4-1} % {revtex4}
\usepackage{amsmath, amssymb}
\usepackage{makecell}
\usepackage{multirow}
\usepackage{CJK}
\usepackage{graphicx}
\usepackage{mathrsfs}
\usepackage{bm}
\usepackage{amsmath}
\usepackage{dcolumn}
\usepackage{epstopdf}
\usepackage{dsfont}
\usepackage{amssymb}
\usepackage{tabularx}
\usepackage{array}
\usepackage{float}
\usepackage{color}
\usepackage{epstopdf}
\usepackage{mathrsfs}

\usepackage[colorlinks, linkcolor=blue,anchorcolor=blue,citecolor=blue,urlcolor=blue]{hyperref}
\usepackage{extarrows}

\begin{document}
\title{Third-order charge transport in a magnetic topological semimetal}
\author{Ziming Zhu}\email{zimingzhu@hunnu.edu.cn}
\affiliation{Key Laboratory of Low-Dimensional Quantum Structures and Quantum Control of Ministry of Education, Department of Physics and Synergetic Innovation Center for Quantum Effects and Applications, Hunan Normal University, Changsha 410081, China}

\author{Huiying Liu}\email{liuhuiying@pku.edu.cn}
\affiliation{Research Laboratory for Quantum Materials, Singapore University of Technology and Design, Singapore 487372, Singapore}

\author{Yongheng Ge}
\affiliation{Key Laboratory of Low-Dimensional Quantum Structures and Quantum Control of Ministry of Education, Department of Physics and Synergetic Innovation Center for Quantum Effects and Applications, Hunan Normal University, Changsha 410081, China}	

\author{Zeying Zhang}
\affiliation{Research Laboratory for Quantum Materials, Singapore University of Technology and Design, Singapore 487372, Singapore}

\author{Weikang Wu}
\affiliation{Research Laboratory for Quantum Materials, Singapore University of Technology and Design, Singapore 487372, Singapore}

\author{Cong Xiao}
\affiliation{Department of Physics, The University of Hong Kong, Hong Kong, China}
\affiliation{HKU-UCAS Joint Institute of Theoretical and Computational Physics at Hong Kong, China}

\author{Shengyuan A. Yang}
\affiliation{Research Laboratory for Quantum Materials, Singapore University of Technology and Design, Singapore 487372, Singapore}
\begin{abstract}
Magnetic topological materials and their physical signatures are a focus of current research. Here, by first-principles calculations and symmetry analysis, we reveal topological semimetal states in an existing antiferromagnet ThMn${_2}$Si$_{2}$. Depending on the N\'{e}el vector orientation, the topological band crossings near the Fermi level form either a double-nodal loop or two pairs of Dirac points, which are all fourfold degenerate and robust under spin-orbit coupling. These topological features produce large Berry connection polarizability, which leads to enhanced nonlinear transport effects. Particularly, we evaluate the third order current response, which dominates the transverse charge current. We show that the nonlinear response can be much more sensitive to topological phase transitions than linear response, which offers a powerful tool for characterizing magnetic topological semimetals.

\end{abstract}

\maketitle
%\end{CJK*}
%\section{Introduction}

Topological semimetals (TSMs) have been attracting tremendous interest in the past decade ~\cite{RevModPhys.88.035005,yanreview2017,RevModPhys.90.015001,RevModPhys.93.025002}.
They are characterized by protected band degeneracies near the Fermi level, which may form a variety of nodal points~\cite{wan2011,murakami2007phase,young2012dirac,fang2012multi,yang2014classification,bradlyn2016beyond,vsmejkal2017electric,yu2022encyclopedia}, nodal lines~\cite{PhysRevLett.113.046401,weng2015topological,fang2016topological}, or nodal surfaces~\cite{liang2016node,zhong2016towards,PhysRevB.96.155105,wu2018nodal} in the momentum space. Due to these degeneracies, the low-energy electron excitations are endowed with exotic characters in the dispersion, pseudospin structure, or topological charge, different from conventional materials.

Currently, despite significant progress in the classifications of TSM states and in the search of these materials~\cite{tang2019efficient,zhang2019catalogue,tang2019comprehensive,vergniory2019complete,xu2020high,jiang2021k,PhysRevB.104.085137,yu2022encyclopedia,PhysRevB.105.085117,PhysRevB.105.104426}, good candidate materials are still quite limited. Here, ``good'' means that
the material should  at least have the band degeneracy  close to the Fermi level and not overlap with other extraneous bands. The challenge is more pronounced regarding the recent research focus of magnetic TSMs ~\cite{wan2011,Xu2011,tang2016dirac,vsmejkal2017electric,hua2018dirac,smejkal2018,shao2019dirac,liu2019magnetic,belopolski2019discovery,PhysRevB.104.165424,Nie2020,bernevig2022progress}. In many magnetic materials, the low-energy bands are complicated owing to the less dispersive $d$ or $f$ bands. In addition, many band degeneracies lose their protection under spin-orbit coupling (SOC) which are often sizable in magnetic materials.

In the meantime, there is urgent need in exploring the physical consequences of TSM states. So far, experimental studies in the field are mainly in terms of spectroscopic probes, linear transport, and magneto-transport. We note that band degeneracies can naturally give rise to strong interband coherence. For example, Weyl points are singularities of Berry curvature~\cite{wan2011}, which is a geometric quantity encoding interband coherence and scales as $\sim 1/(\Delta \varepsilon)^2$ in terms of the energy splitting $\Delta \varepsilon$ between two bands. Indeed, this underlies the large anomalous Hall response~\cite{PhysRevB.84.075129,PhysRevLett.113.187202} and the chiral anomaly~\cite{PhysRevD.22.3080,PhysRevB.88.104412} effect proposed for Weyl semimetals. Following this thought, one naturally wonders whether there are other band geometric quantities enhanced in TSMs and what physical effects they may bring about.

In this work, we first reveal a high-quality magnetic TSM state in an existing antiferromagnetic (AFM) material ThMn${_2}$Si$_{2}$. We show that depending on its N\'{e}el vector direction, the band degeneracies can form a double-nodal loop or two pairs of Dirac points close to the Fermi level. These magnetic band crossings are fourfold degenerate and robust against SOC. Using this material as an example, we show that TSMs feature strongly enhanced Berry connection polarizability (BCP)~\cite{PhysRevLett.112.166601,PhysRevB.91.214405,gao2019semiclassical,PhysRevB.105.045118}, which is an intrinsic band geometric quantity and scales as $\sim 1/(\Delta \varepsilon)^3$. In ThMn${_2}$Si$_{2}$, it leads to a pronounced third-order charge current response to a driving $E$ field, which dominates the transverse current transport. Furthermore, we show that the third-order signal is much more sensitive to the change in band topology than the linear order, hence it offers a promising tool for characterizing TSMs.

\begin{figure*}[t!]
\centerline{\includegraphics[width = 0.9\linewidth]{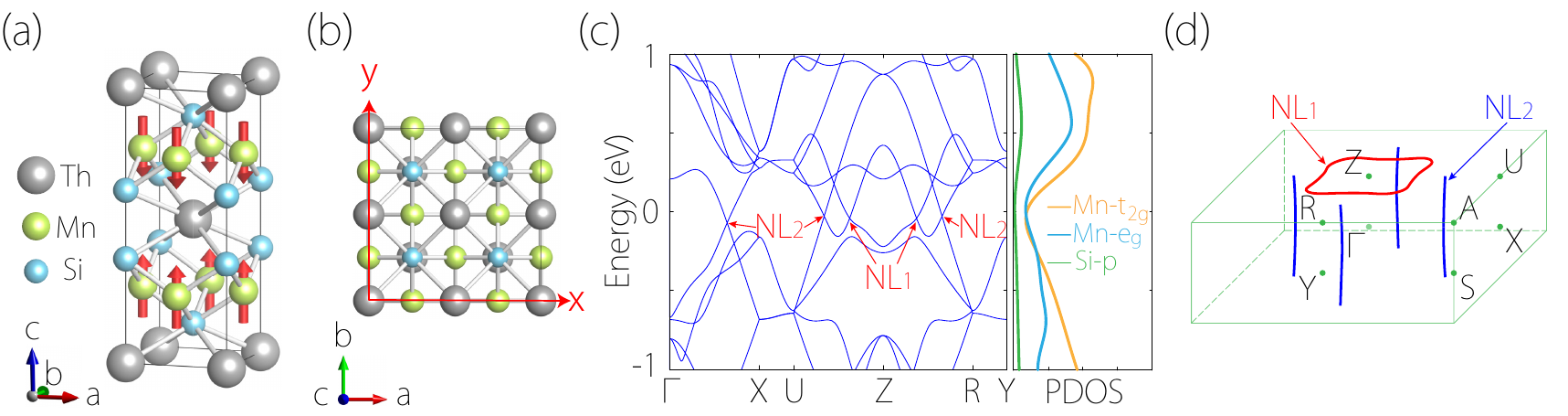}}
\caption{\label{fig1} (a) Perspective view and (b) top view of ThMn$_2$Si$_2$ lattice.  The magnetic structure in (a) represents the AFM-$z$ state. (c) Band structure of ThMn$_2$Si$_2$ along with PDOS in the absence of SOC. The arrows mark the band crossing points belonging to two types of nodal lines, NL$_1$ and NL$_2$, as schematically illustrated in (d).}
\end{figure*}

{\color{blue}\textit{ThMn$_2$Si$_2$: structure and magnetism.}} The ThMn$_2$Si$_2$ single crystal was first synthesized in the 1960s~\cite{sikirica1964thorium,ban1965crystal}. It has the tetragonal CeAl$_2$Ge$_2$-type structure with space group $I_4/mmm$ (No.~139). As shown in Fig.~\ref{fig1}, the structure consists of atomic layers stacked along the $c$-axis ($z$ direction), with the Mn layer separated by the Si-Th-Si sandwiches.
The Th atoms occupy the 2\emph{a} Wyckoff position, whereas Mn (Si) atoms are at the position of 4\emph{d} (4\emph{e}).
For the conventional cell in Fig.~\ref{fig1}, the experimental lattice constants are $a=b=4.021$ {\AA} and $c=10.493$ {\AA}~\cite{ban1965crystal}. These values are adopted in our first-principles calculations.

%The compound contains the following generators: four-fold rotation $C_{4z}$, inversion $\mathcal{P}$ and twofold rotation $C_{2x}$,
%and  $\tilde{E}=\{E|\frac{1}{2}a,\frac{1}{2}a,\frac{1}{2}c\}$ with E being the identity.
%Combining inversion with the twofold rotations, we also have three mirror symmetries $M_x$, $M_y$, and $\tilde{M}_z$, where
%$\tilde{M}_z=\{M_z|\frac{1}{2}a,\frac{1}{2}a,\frac{1}{2}c\}$ is a glide mirror.

Early magnetic measurements showed that ThMn$ _{2} $Si$ _{2}$ is AFM with a N\'{e}el temperature $\sim 483$ K~\cite{ban1975magnetic}. From neutron diffraction result, the magnetism is mainly from Mn and has the A-type configuration as illustrated in Fig.~\ref{fig1}(b), i.e., the coupling is ferromagnetic within each Mn layer and is AFM between the layers. The magnetic easy axis is along $c$. The magnetic moment at Mn site was found to be $\sim 1.75 \mu_B$ at 78 K~\cite{ban1975magnetic} and $<2\mu_B$ at 4.2 K~\cite{narasimhan1976magnetic}. All these features have been successfully reproduced by our first-principles calculations. 
%(see Supplemental Material for details).

\begin{figure}[htp!]
	{\includegraphics[clip,width=8.2cm]{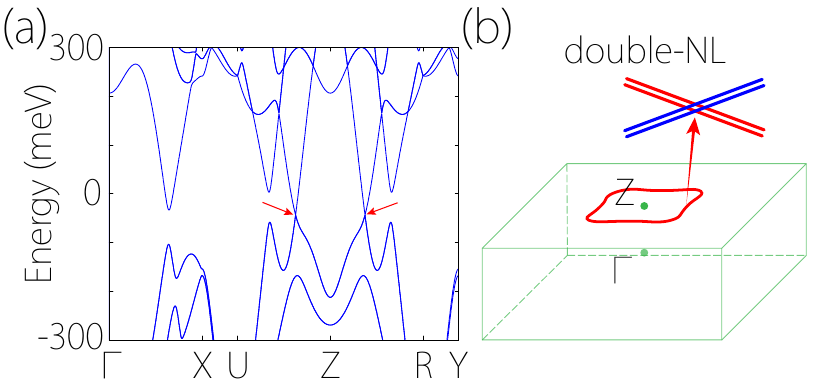}}
	%magstru_3.png
	\caption{\label{fig2}
(a) Band structure with SOC for the AFM-$z$ state. The arrows indicate the crossing points on the double-nodal loop, as illustrated in (b). Each point is formed by the crossing of four bands. }
\end{figure}

{\color{blue}\textit{Nodal lines in the absence of SOC.}} Let's first consider the band structure of AFM ThMn$ _{2}$Si$ _{2}$ in the absence of SOC. As shown in Fig.~\ref{fig1}(c), the system exhibits a semimetal character. The
low-energy bands around Fermi level are mainly from  Mn-3\textit{d} and Si-2\textit{p} orbitals. One observes multiple band crossings close to the Fermi level~\footnote{Here, if counting spin, each band is doubly degenerate, and the crossing is fourfold degenerate.}. A careful scan over the Brillouin zone (BZ) shows that they belong to two families of nodal lines, which form different winding patterns over the BZ torus~\cite{li2017type,vanderbilt2018berry}. As illustrated in Fig.~\ref{fig1}(d), the first family (denoted as NL$_1$) consists of a single nodal ring located within the $k_z=\pi$ plane and centered at $Z$, whereas the second (denoted as NL$_2$) includes four nodal lines in the $k_x=0$ and $k_y=0$ planes, each traversing the BZ in the $z$ direction.

To understand the protection of these nodal lines, we note that without SOC, the two spin channels are decoupled hence can be analyzed separately. Each spin channel can be regarded as spinless and has an \emph{effective} time reversal symmetry $T$. In addition, each channel has the crystal symmetry of inversion ($\mathcal{P}$), vertical mirrors $M_x$ and $M_y$, and horizontal glide mirror $\tilde{M}_z=\{M_z|\frac{1}{2}\frac{1}{2}0\}$ (at a Mn layer). Therefore, the nodal lines here actually enjoy a double protection. First, in each spin channel, the spinless $\mathcal{P}T$ symmetry dictates a quantized $\pi$ Berry phase on a small loop encircling a line, protecting it from opening a gap. Second, each line is protected by a mirror symmetry as the two crossing bands have opposite mirror eigenvalues. This also constrains the nodal lines in the three mirror-invariant  planes of the BZ.

{\color{blue}\textit{Double-nodal loop and Dirac points under SOC.}} Now, we study the band structure when SOC is included. We first consider the ground state with the N\'{e}el vector along the $z$ direction, denoted as the AFM-$z$ state. The system has a magnetic space group of
$P_{I}{4/nnc}$. The calculated band structure in Fig.~\ref{fig2}(a) overall looks similar to Fig.~\ref{fig1}(c). Focusing on the nodal lines in Fig.~\ref{fig2}(a), one observes that the NL$_2$ lines are gapped out, however, surprisingly, the NL$_1$ ring is still maintained under SOC.

With SOC, each state $|u\rangle$ in Fig.~\ref{fig2} is degenerate with a partner $\mathcal{PT}|u\rangle$, where $\mathcal{T}$ is the genuine time reversal operation and $\mathcal{P}$ here acts on central Th site in Fig.~\ref{fig1}(a). Hence, the NL$_1$ ring is fourfold degenerate, formed by crossing between two doubly $\mathcal{PT}$-degenerate bands. This is made possible if the  $\mathcal{PT}$ partners share the same $\tilde{M}_z$ eigenvalue and the crossing is between bands with opposite $\tilde{M}_z$ eigenvalues, as illustrated in Fig.~\ref{fig2}(b).

To verify this, note that
\begin{equation}
  \tilde{M}_z^2=-t_{110}=-e^{-ik_x-ik_y},
\end{equation}
with $t_{110}$ the translation of one lattice unit respectively in $x$ and $y$ direction, so its eigenvalues are given by
$g_z=\pm i e^{-i(k_x+k_y)/2}$.
Due to the offset between the inversion center (Th site) and the mirror plane (Mn layer), the commutation relation between $\mathcal{P}$ and $\tilde{M}_z$ is
$\tilde{M}_z\mathcal{P}=t_{111}\mathcal{P}\tilde{M}_z$.
It follows that for any state $|u\rangle$ with $\tilde{M}_z|u\rangle=g_z|u\rangle$, its partner satisfies
\begin{equation}
   \tilde{M}_z(\mathcal{PT}|u\rangle)=-e^{-ik_z}g_z (\mathcal{PT}|u\rangle).
\end{equation}
Thus, in the $k_z=\pi$ plane, the $\mathcal{PT}$ partners indeed share the same eigenvalue $g_z$.

This kind of fourfold nodal loop robust under SOC was initially proposed by Fang \emph{et al.} ~\cite{fang2012multi} and was named the double-nodal loop. Its material realizations are very rare, especially in magnetic systems, being only predicted in MnPd$_2$ till now ~\cite{shao2019dirac}.

\begin{figure}[htp!]
	{\includegraphics[clip,width=8.2cm]{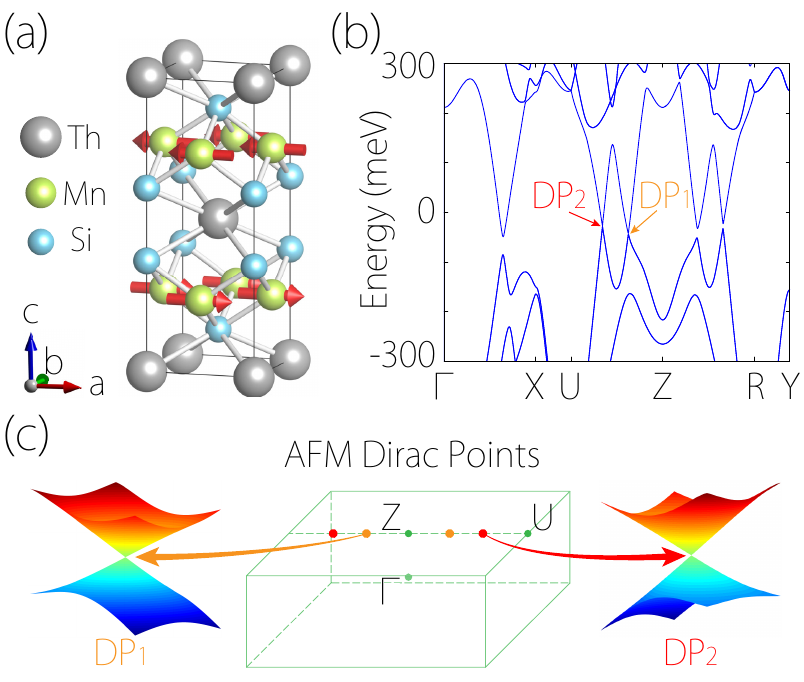}}
	%magstru_3.png
	\caption{\label{fig3}  (a) The magnetic structure for the AFM-$x$ state. (b) shows the corresponding band structure with SOC. The arrows mark two (of the four) Dirac points on the $Z$-$U$ path. (c) illustrates the distribution and the dispersion of these Dirac points.}
\end{figure}

A unique feature of magnetic TSMs is that their topological states can be controlled by rotating the magnetic order parameter, e.g., by spin torques, applied field or strain, which changes the symmetry of the system. Here, let's consider the case when the N\'{e}el vector is along the $x$ direction (referred to as AFM-$x$ state). This breaks the fourfold rotation along $z$ and the magnetic space group becomes $P_{I}{mmn}$. As shown in Fig.~\ref{fig3}, points on the NL$_1$ and NL$_2$ nodal lines are almost all gapped out except for four points, namely the intersection points of these lines with the $Z$-$U$ path. These points represent fourfold degenerate AFM Dirac points. Hence, when the N\'{e}el vector rotates from $z$ to $x$ direction, the system transitions from
a magnetic double-nodal loop semimetal to an AFM Dirac semimetal.

To understand this topological phase transition, we note that: (1) the $\mathcal{PT}$ symmetry is maintained regardless of the N\'{e}el vector direction; (2) $\tilde{M}_z$ is broken for AFM-$x$, so the double-nodal loop is no longer protected; (3) there emerges a twofold screw axis $\tilde{C}_{2x}=\{C_{2x}|\frac{1}{2}\frac{1}{2}0\}$ as indicated in Fig.~\ref{fig3}(a). This symmetry, together with $\mathcal{PT}$, protects the four Dirac points on the $\tilde{C}_{2x}$-invariant path $Z$-$U$.
The analysis is similar to that for the AFM-$z$ case hence is relegated to Supplemental Material.

\begin{figure}[t!]
	{\includegraphics[clip,width=8.2cm]{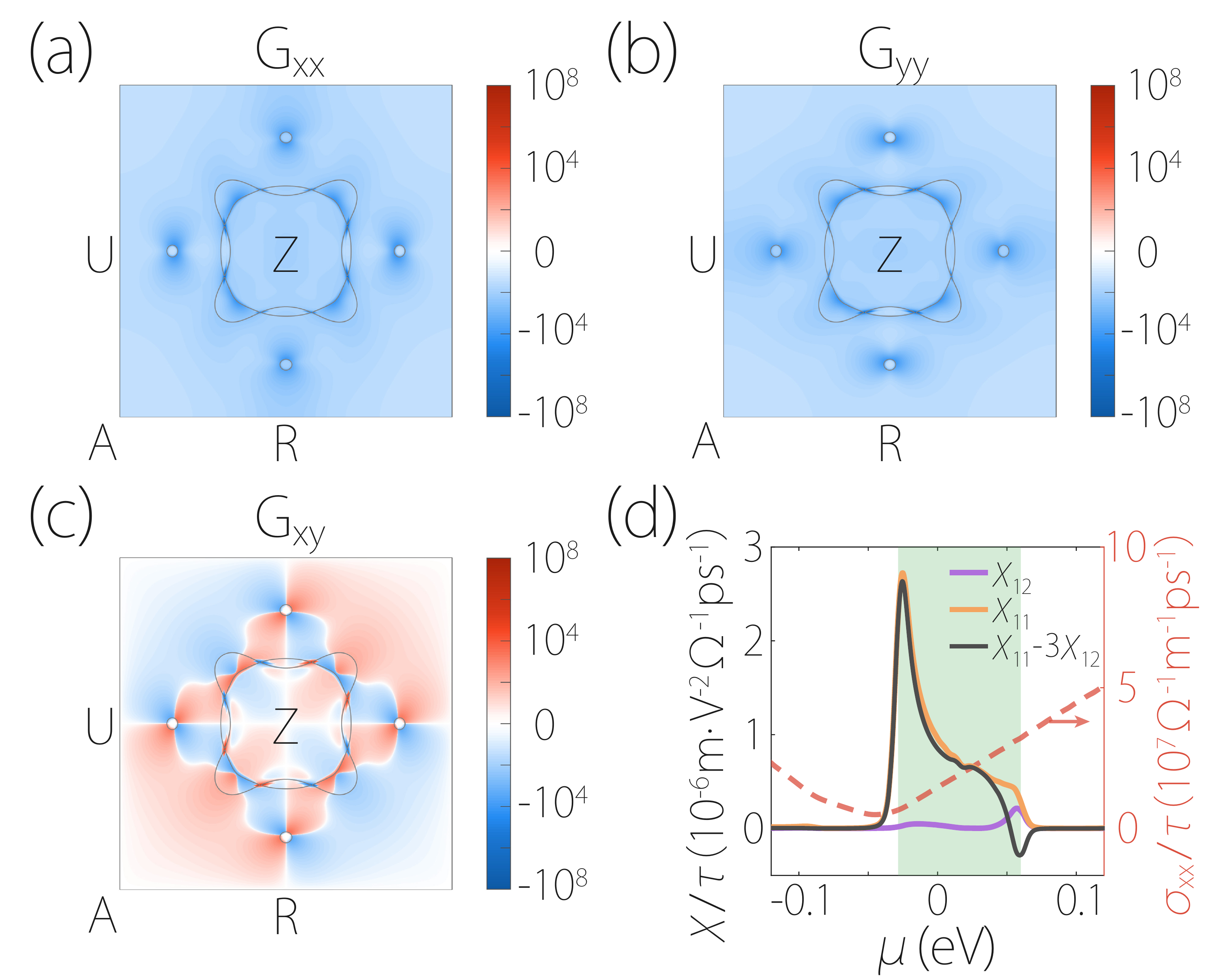}}
	%magstru_3.png
	\caption{\label{fig4} (a-c) Calculated distribution of BCP tensor elements in the $k_{z}=\pi$ plane  for the AFM-$z$ state. The grey lines depict the Fermi surface. The unit of $G_{ab}$ is \AA$^{2}\cdot$V$^{-1}$. (d) Third-order conductivity tensor elements (divided by $\tau$) versus the chemical potential $\mu$ for the AFM-$z$ state. Here, we also plot the longitudal conductivity $\sigma_{xx}$ (red dashed line, right axis). The energy range of the double-nodal loop is indicated by the green shaded region.}
\end{figure}

{\color{blue}\textit{BCP and third-order current response.}} Like the Berry curvature, BCP is an intrinsic band geometric quantity, which characterizes the positional shift of Bloch electrons under an applied $E$ field~\cite{PhysRevLett.112.166601,PhysRevB.91.214405,gao2019semiclassical,PhysRevB.105.045118}. It is a second-rank tensor, and for a band $n$, it can be expressed as (we set $e=\hbar=1$)
\begin{equation}
G_{ab}(\bm k)=2\text{Re}\sum_{m\neq n}\frac{(v_{a})_{nm}(v_{b})_{mn}}{(\varepsilon_{n}-\varepsilon_{m})^3},\label{eq:BCP}
\end{equation}
where $a$ and $b$ label the Cartesian components, the $v$'s are interband velocity matrix elements, and $\varepsilon_n(\bm k)$ is the band energy. From  (\ref{eq:BCP}), one can see that BCP $\sim 1/(\Delta \varepsilon)^3$, so it should be strongly enhanced at small gap regions, especially the band degeneracies. In Fig.~\ref{fig4}, we plot the BCP components for the AFM-$z$ state in the $k_z=\pi$ plane where the double-nodal loop is located. One clearly observes that BCP is concentrated around the nodal loop as well as the four small-gap spots where the original NL$_2$ lines cross the plane.

In TSMs, band degeneracies exist near the Fermi level, which means pronounced BCP would appear for the low-energy states, thereby impacting physical properties of the system. It was recently shown that BCP underlies many nonlinear response properties of solids~\cite{PhysRevB.105.045118,gao2019semiclassical,lai2021third,PhysRevLett.127.277201,PhysRevLett.127.277202,PhysRevLett.129.086602}. Here, we shall consider the third-order charge current response. This is because the ground state, i.e., AFM-$z$ state, of ThMn$_2$Si$_2$ has a quite high symmetry: It has the symmetry $\mathcal{T}t_{00\frac{1}{2}}$, the inversion $\mathcal{P}$, and fourfold rotation $C_{4z}$. Considering in-plane transport, i.e., applied $E$ field and response current are in the $x$-$y$ plane, these symmetries suppress both the linear- and second-order current responses in the direction transverse to $E$. Thus, the third-order current $j\sim E^3$ will be the leading order transverse response.

In the extended semiclassical theory, BCP determines the third-order current that is linear in the electron relaxation time $\tau$~\cite{PhysRevB.105.045118}. Specifically, the corresponding third-order conductivity tensor can be expressed as~\cite{PhysRevB.105.045118}
\begin{align}
\chi_{abcd}& = \tau\Big[\int[d\bm{k}](-\partial_{a}\partial_{b}G_{cd}+\partial_{a}\partial_{d}G_{bc}-\partial_{b}\partial_{d}G_{ac})f_{0}\nonumber \\
 & \qquad +\frac{1}{2}\int[d\bm{k}]v_{a}v_{b}G_{cd}f_{0}''\Big],\label{eq:chiI}
\end{align}
where $[d\bm{k}]$ stands for $\sum_n d\bm k/(2\pi)^d$ with $d$ the dimension of the system, the $v$'s here are the intraband velocity for the band $n$, $\partial_{a} \equiv \partial_{k_{a}}$, and $f_0$ is the Fermi distribution function. Considering in-plane transport, the indices $a,b,c,d\in\{x,y\}$. Obviously, the $\chi$ tensor is most easily evaluated in a coordinate system adapted to the crystal, as in Fig.~\ref{fig4}. The direction of applied $E$ field can be specified by its polar angle $\theta$, i.e., $\bm E=E(\cos\theta, \sin\theta,0)$. Note that the expression (\ref{eq:chiI}) includes both the longitudinal and transverse responses. Focusing on the transverse third-order current $j_\bot^{(3)}$ which is along $\hat{z}\times \bm E$, we can write $j_\bot^{(3)}=\chi_\bot(\theta) E^3$ in terms of a third-order transverse conductivity $\chi_\bot$. For the ground-state ThMn$_2$Si$_2$ with $C_{4z}$ symmetry, we find
\begin{equation}
  \chi_{\perp}(\theta)=-\frac{1}{4}(\chi_{11}-3\chi_{12})\sin4\theta,\label{eq:trans_C4}
\end{equation}
where $\chi_{11}=\chi_{xxxx}$, and $\chi_{12}=(\chi_{xxyy}+\chi_{xyxy}+\chi_{xyyx})/3$.

In Fig.~\ref{fig4}(d), we plot the involved tensor elements for ThMn$_2$Si$_2$ as a function of chemical potential. One observes that the response is peaked around the intrinsic Fermi level, in a window overlaps with the energy range of the double-nodal loop (the green shaded region).
This confirms our claim that in TSMs, band degeneracies  tend to enhance BCP and generate pronounced nonlinear effects.
%In Fig., we plot the angular dependence of the transverse conductivity, which can be tested in experiment using a multiple-lead geometry.

{\color{blue}\textit{Probing topological phase transition.}}
We have shown that third-order response is closely connected to the band degeneracies in TSMs. As a result, when there is a change in the degeneracy, i.e., when the system undergoes a topological phase transition, a significant change in the nonlinear response can be expected.

\begin{figure}[t!]
	{\includegraphics[clip,width=8.2cm]{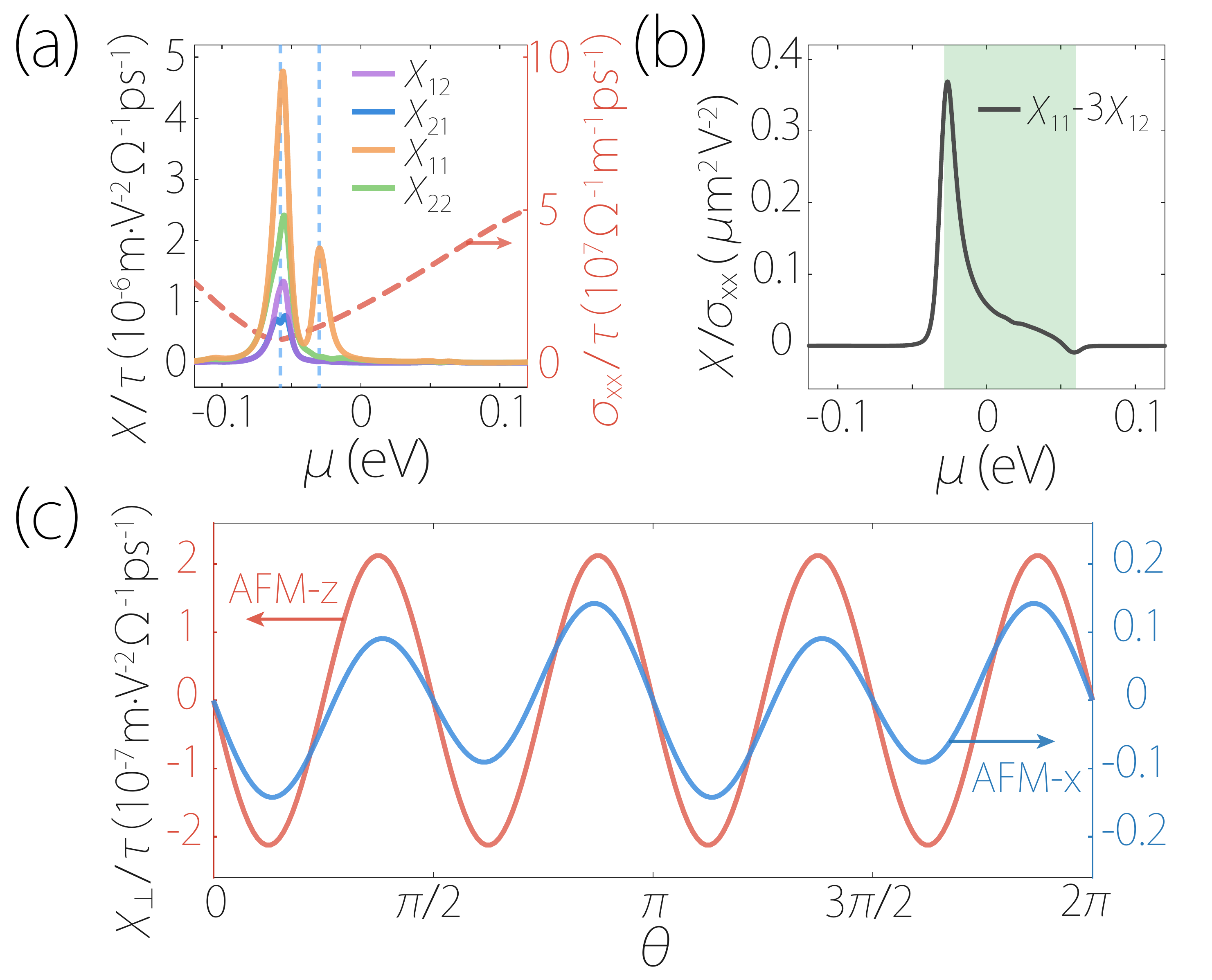}}
	%magstru_3.png
	\caption{\label{fig5} (a) Third-order conductivity tensor elements (divided by $\tau$) versus $\mu$ for the AFM-$x$ state. The two vertical dashed lines mark the energies of the AFM Dirac points.
The corresponding longitudal conductivity $\sigma_{xx}$ is also shown by the red dashed curve. (b) The ratio $(\chi_{11}-3\chi_{12})/\sigma_{xx}$ plotted versus $\mu$ for the AFM-$z$ state. (c) The comparison of the third-order transverse conductivity $\chi_{\perp}(\theta)$ (divided by $\tau$) for AFM-$z$ state (red line) and AFM-$x$ state (blue line). The AFM-$z$ result is an order of magnitude larger than AFM-$x$, and they show different angular dependence.}
\end{figure}

Here, we demonstrate this point in ThMn$_2$Si$_2$. When the N\'{e}el vector direction rotates from $z$ to $x$, a topological phase transition happens, with the original degeneracy at the double-nodal loop replaced by four Dirac points. Note that for the AFM-$x$ state, the $C_{4z}$ symmetry is broken, so there are more independent elements of $\chi$, and the expression for $\chi_\bot$ changes to
\begin{align}
 \begin{split}
\chi_{\perp}(\theta) & = (3\chi_{21}-\chi_{11})\cos^{3}\theta\sin\theta \\
 & \qquad +(\chi_{22}-3\chi_{12})\cos\theta\sin^{3}\theta.\label{eq:trans}
 \end{split}
\end{align}
where  $\chi_{22}=\chi_{yyyy}$,
and $\chi_{21}=(\chi_{yyxx}+\chi_{yxyx}+\chi_{yxxy})/3$. Fig.~\ref{fig5}(a) plots the relevant tensor elements versus chemical potential. Compared with Fig.~\ref{fig4}(d), one can see a dramatic change in the response. The contribution from the original nodal loop is largely suppressed. Instead, the two peaks in Fig.~\ref{fig5}(a) are perfectly aligned with the energies of the two pairs of Dirac points (as marked by the two vertical dashed lines).

In Fig.~\ref{fig5}(c), we plot $\chi_\bot(\theta)$ for the two states in the same figure, as a function of angle $\theta$. One can see that the nonlinear response for AFM-$x$ state is reduced by an order of magnitude, despite the overall similarity of the two band structures (see Fig.~\ref{fig2}(a) and Fig.~\ref{fig3}(b)). For comparison, in Fig.~\ref{fig4}(d) and Fig.~\ref{fig5}(a), we also plot the linear Drude conductivity for each state, which shows much less change between the two states.
This demonstrates that the nonlinear response from the BCP is indeed very sensitive to the change in band topology, thereby offering a promising tool for probing topological phase transitions.

In addition, from Fig.~\ref{fig5}(c), one notes that due to the different symmetry, the third-order response $\chi_\bot(\theta)$ exhibits different angular dependence for the two states. In the AFM-$z$ state, $\chi_\bot(\theta)$ has a period of $\pi/2$, whereas the period is doubled for AFM-$x$. This feature can be tested in experiment using a multiple-lead geometry.

{\color{blue}\textit{Discussion.}} We have revealed ThMn$_2$Si$_2$ as a high-quality magnetic TSM. Its magnetism persists at room temperature, and its band structure is clean with degeneracies close to the Fermi level. By controlling the N\'{e}el vector direction, the system realizes an magnetic double-nodal loop semimetal or a AFM Dirac semimetal. These topological band features can be readily probed by angle-resolved photoemission spectroscopy (ARPES)~\cite{RevModPhys.93.025002}.

We show that TSMs can host enhanced BCP and propose the third-order current response as a sensitive tool for characterizing TSMs. Experimentally, the nonlinear signal is typically detected using the lock-in technique with a low-frequency ac driving field, which was successfully applied in several recent experiments on nonlinear transport\cite{kang2019electrically,ma2019observation,lai2021third,wang2022room}.

As discussed, the third-order conductivity $\chi_\bot$ here is closely connected with BCP, which is peaked at the band degeneracies. In TSMs, this peak occurs near the intrinsic Fermi level, as shown in Fig.~\ref{fig4}(d). Meanwhile, the linear longitudinal conductivity $\sigma$ typically reaches a minimum at the same range, because as a semimetal the density of states is suppressed there, as confirmed in Fig.~\ref{fig4}(d) and Fig.~\ref{fig5}(a).
This contrasting behavior again manifests the important geometric origin of nonlinear transport. In Fig.~\ref{fig5}(b), we further plot the ratio $\chi/\sigma$ versus the chemical potential, which is even more enhanced at topological band degeneracies and serves a band intrinsic quantity free of $\tau$ that can be compared between different materials.

%\appendix

%\begin{acknowledgements}
%	The authors thank Zeying Zhang, Quan-Sheng Wu and D. L. Deng for valuable discussions. This work is supported by the National Natural Science Foundation of China (NSFC) (Grant No.~11704117 and 11974076) and the Singapore MOE AcRF Tier 2 (MOE-T2EP50220-0011). We acknowledge computational support from Texas Advanced Computing Center and H2 clusters in Xi'an Jiaotong University
%\end{acknowledgements}

\bibliography{afm_refs}

\bibliographystyle{apsrev4-1}

\end{document}